# ZnO, ZnMnO and ZnCoO films grown by Atomic Layer Deposition

M.I. Łukasiewicz[1], A. Wójcik-Głodowska[1], E. Guziewicz[1], A. Wolska[1], M. T. Klepka[1],

P. Dłużewski[1], R. Jakieła[1], E. Łusakowska[1], K. Kopalko[1], W. Paszkowicz[1], Ł. Wachnicki[1],

B.S. Witkowski[1], W. Lisowski[2], M. Krawczyk[2], J.W. Sobczak[2], A. Jabłoński[2],

M. Godlewski[1,3]

[1] Institute of Physics Polish Academy of Sciences, Al. Lotników 32/46,

02-668 Warsaw, Poland

[2] Institute of Physical Chemistry, Polish Academy of Sciences, Kasprzaka 44/52,

01-224 Warsaw, Poland

[3] Dept. Mathematics and Natural Sciences, College of Sciences UKSW, Dewajtis 5,

01-815 Warsaw, Poland

**Abstract**

Despite many efforts the origin of a ferromagnetic (FM) response in ZnMnO and ZnCoO is still not clear. Magnetic investigations of our samples, not discussed here, show that the room temperature FM response is observed only in alloys with a non-uniform Mn or Co distribution. Thus, the control of their distribution is crucial for explanation of contradicted magnetic properties of ZnCoO and ZnMnO reported till now. In the present review we discuss advantages of the Atomic Layer Deposition (ALD) growth method, which enables us to control uniformity of ZnMnO and ZnCoO alloys. Properties of ZnO, ZnMnO and ZnCoO films grown by the ALD are discussed.

# 1. Introduction

Alloys of ZnO with transition metal ions (TM), such as manganese (ZnMnO) or cobalt (ZnCoO), are intensively investigated for possible spintronic applications. For these applications ZnTMO alloys should show a ferromagnetic (FM) response at room temperature (RT), which was predicted by the theory for p-type ZnMnO [1]. Moreover, we should achieve spin polarization of free carries at RT. Unfortunately, this is not possible if the FM response is due to some "local" properties (inclusions or accumulations) and not to "volume" properties of the alloys studied.

At first spintronic applications of ZnTMO alloys looked very likely. The RT FM response has been reported in several cases, not only for ZnMnO and ZnCoO (see e.g. [2-24]), but also for many other GaN and ZnO based diluted magnetic semiconductors (DMSs). However, the RT FM response reported for ZnMnO [4-15] or ZnCoO [16-24] is most likely due to inclusions of foreign phases of various TM oxides, defects, or due to TM accumulations, and not to "volume" properties of the alloys studied. Not surprisingly, the information on samples uniformity is crucial. In many cases it allows to account for the magnetic properties of investigated samples.

Uniformity of TM alloys is affected by two effects. First, it is affected by large diffusion coefficients of TM ions and a strong driving force pushing TM ions to group together in small nanoclusters. Secondly, by the fact, that TM ions readily form different TM oxides. Thus, it is difficult to eliminate inclusions of foreign phases in ZnTMO alloys.

The first effect was discussed by Schilfgaarde and coworkers [25,26]. Following the results of their calculations, we expected that deposition of uniform DMS alloys is quite challenging. In particular, it is difficult in high temperature growth processes. For example, a strong non-uniformity of Co distribution was observed by us in electron spin resonance (ESR) investigations of bulk ZnO:Co [27]. Even though the crystals investigated by us contained a relatively low concentration of Co ions (on an impurity level) we detected ESR signals of Co-Co pairs [27]. Such signals were not expected in samples with a statistical (random) Co distribution. Not surprisingly, problems with uniformity of TM distribution were reported for most of ZnTMO samples grown by quite different growth techniques. In the consequence, the often reported FM response may entirely relate to this fact. Moreover, the problems with alloys uniformity likely explain a low reproducibility of the data of magnetic investigations, as reported in some of the references (see e.g. [16,17]).

In the present work we review our results for two wide band gap oxides, ZnMnO and ZnCoO, grown by the technique of Atomic Layer Deposition (ALD). We discuss the methods of deposition of uniform ZnMnO and ZnCoO films, based on the results of our recent investigations [15,18,24,28,29]. This allows us to clarify the origin of FM responses, which were observed by us only for non-uniform samples, i.e. samples with a non-uniform TM distribution. Based on these observations, the FM responses are related to inclusions of foreign phases (in ZnMnO) and metal accumulations (in ZnCoO). Results for ZnTMO layers are compared to the ones we obtained for ZnO films, studied by us separately. For example, growth modes (orientations of the ALD-grown ZnO, ZnMnO and ZnCoO samples) are discussed. This information is important to account for very contradicting reports on strength of magnetic interactions in e.g. ZnCoO [30].

The paper is organized in the following way. First, experimental methods used are described. Then we discuss ZnO deposition by the ALD and selection of precursors for ZnO, ZnMnO and ZnCoO deposition. This is followed by the discussion of growth modes, influence of a purging time and of growth temperature. Finally, we describe how to avoid non-uniformity of Mn/Co distribution by selecting the appropriate ratios of the ALD pulses. We also explain why magnetic response detected may depend on a sample thickness.

## 2. Experimental methods

Structural properties of ZnO and ZnTMO films were characterized by X-ray diffraction (XRD) in a full angular range using the Panalytical X'Pert MRD diffractometer. Results of the XRD investigations are used not only to verify deposition of ZnO and ZnTMO, but also to trace the growth modes (see discussion given below). The XRD data allow also to evaluate quality of our samples and size of crystallites (from a Full Width at Half Maximum (FWHM) of the relevant XRD peaks), as shortly discussed later on.

The X-ray absorption measurements (XANES and EXAFS) were performed at the K edge of Mn and Co at DESY- Hasylab (Cemo and A1 stations). Measurements were performed at liquid nitrogen temperature using a 7-element silicon fluorescence detector. XANES and EXAFS study allows us to confirm Mn/Co substitution to ZnO, but also allows detection of some inclusions of foreign phases (Mn oxides and Co metal).

Composition investigations were performed with the Secondary Ion Mass Spectroscopy (SIMS) technique using a CAMECA IMS6F micro-analyzer. Magnetization investigations (we do not discuss these results in the present work) were done using a home-made SQUID (Superconducting Quantum Interference Device) magnetometer and the commercial Bruker 300 X-band electron spin resonance (ESR) spectrometer.

The surface morphology was investigated with the Dimension Icon Atomic Force Microscope (AFM) of the Bruker company. Scanning electron microscopy (SEM) and cathodoluminescence (CL) investigations were performed with the Hitachi SU-70 microscope equipped with the GATAN MONO CL3 CL system. CL spectra were taken at relatively low accelerating voltages either at RT or at 5 K. We performed CL depth profiling and collected maps of depth and in-plane variations of the CL intensity.

The XPS spectra were recorded with PHI 5000 VersaProbeTM scanning ESCA Microprobe using monochromatic Al-K radiation (hν = 1486.6 eV). The high-resolution XPS spectra were collected with the analyzer pass energy of 23.5 eV and the energy step of 0.1 eV. Shirley background subtraction and peak fitting with Gaussian-Lorentzian-shaped profiles were performed to analyze the high-resolution XPS spectra. Depth-profiling XPS investigations were performed by removing layer after layer of the sample by sputtering. The first 15 nm of the etched film was removed using 0.5 kV Ar ion etching, with the rate of 1.5 nm per minute, and then the sputter rate of 15 nm per minute was used (2 kV Ar ion etching).

High resolution transmission electron microscopy (HRTEM) studies were performed at 300 kV electron beam energy with a Titan 80-300 Cubed Cs image corrected microscope equipped with an energy dispersion X-ray (EDX) spectrometer allowing a chemical analysis. The cross sectional specimens have been prepared by ion milling proceeded with a mechanical dimpling.

Film thickness was controlled from interference patterns with the Nanocal2000. The results of these measurements were then verified by the SIMS, α-step and SEM (cross-section images) investigations.

## 3. Experimental results and their discussion

### 3.1 Growth of ZnO and ZnTMO alloys by the ALD

Below we describe the methods of deposition of ZnO layers using the ALD and only then, after setting the growth conditions of ZnO, of uniform alloys of ZnTMO (TM stands for Mn and Co). By non-uniform ZnTMO films we mean here the ones showing TM accumulations, inclusions of foreign phases (TM oxides), and TM metal accumulations. We discuss correlations between the growth conditions and the properties of samples.

ZnO, ZnMnO and ZnCoO samples discussed in the present work were grown using either the F-120 ALD reactor produced by the "Microchemistry" company or the Savannah-100 ALD reactor by the Cambridge NanoTech. We used several types of organic zinc, manganese and cobalt precursors selected for the one purpose – a reduction of the growth temperature. We assumed that a low growth temperature is a critical condition for deposition of uniform ZnTMO alloys. This assumption turned out to be true and deposition at low temperature (LT) of uniform ZnMnO and ZnCoO layers allowed us to control magnetic responses of investigated materials, as discussed below in more details.

#### 3.1.1 Zinc acetate as zinc precursor

The first zinc precursor used by us was zinc acetate ($Zn(CH_3COO)_2$), successfully used by us for a growth of ZnO nanoparticles by a hydrothermal method. We tested suitability of this precursor for deposition of ZnO films and only then, after setting the optimal growth conditions, we used zinc acetate for deposition of ZnMnO.

Layers of ZnO and ZnMnO were grown at LT by a double exchange chemical reaction:

$$Zn(CH_3COO)_2 + H_2O \rightarrow ZnO + 2CH_3COOH$$

with $CH_3COOH$ being the reaction product in a gaseous form. This product was removed by purging cycles. In all processes discussed in the present review (with zinc acetate, and then with diethylzinc (DEZn) and dimethylzinc (DMZn) used as zinc precursors) we used a high purity nitrogen gas as a purging gas and water vapors (deionized water) as an oxygen precursor.

It turned out that zinc acetate works also as a mono-precursor. McAdie [31] described decomposition of zinc acetate, occurring in nitrogen atmosphere at increased temperature. The two alternative decomposition routes described in the above work are as follows:

$Zn(CH_3COO)_2$ (solid) $\rightarrow$ $Zn(CH_3COO)_2$ (liquid) $\rightarrow$ ZnO (solid) $\rightarrow$ $CH_3COCH_3$ (gas) + $CO_2$ (gas)

or

$4Zn(CH_3COO)_2$ (solid) $\rightarrow$ $Zn_4O(CH_3COO)_6$ (solid) + $(CH_3CO)_2O$ (gas) $\rightarrow$ $Zn_4O(CH_3COO)_6$ (liquid) $\rightarrow$ 4ZnO (solid) + $3CH_3COCH_3$ (gas) + $3CO_2$ (gas)

In both these cases ZnO is produced as a decomposition product. Consequently, only at low temperature (below 300 ºC) we needed oxygen precursor to grow ZnO.

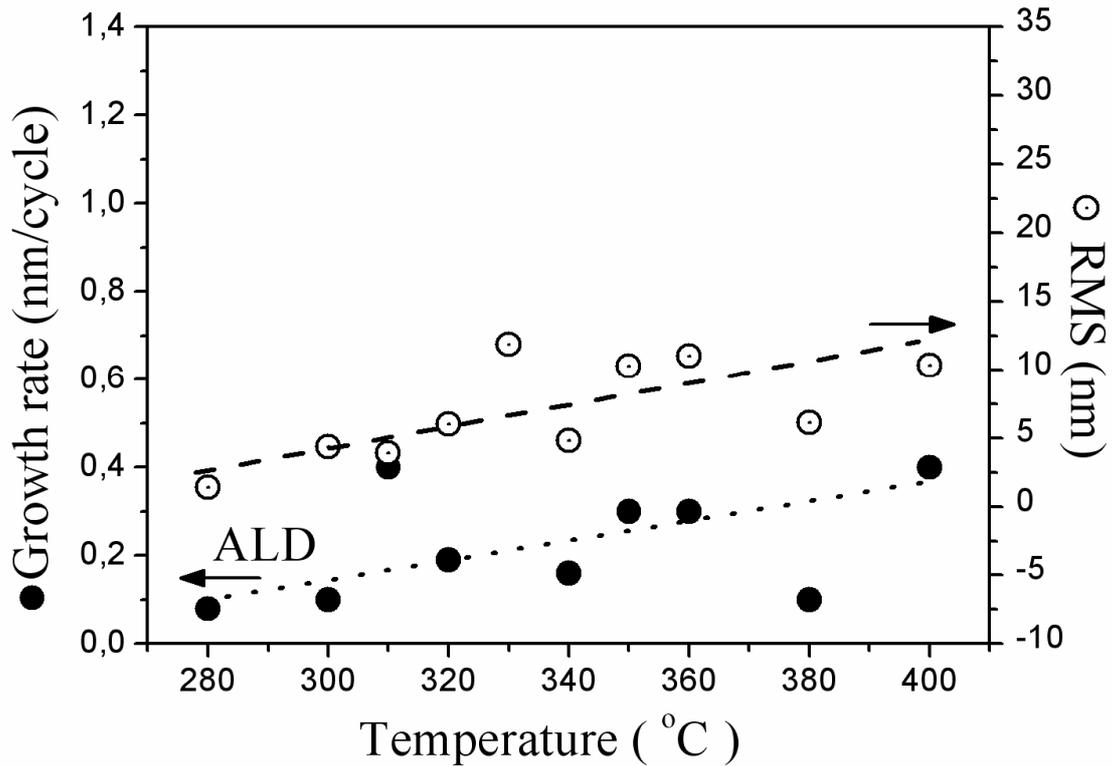

**Figure 1.** Temperature dependence (temperature of a substrate is given) of the growth rates and the mean roughness (RMS) of ZnO layers grown by the ALD with a zinc acetate and deionized water vapors. The RMS was determined from the AFM scans taken from 10x10 µm regions.

For an increased temperature we always observed a mixture of the ALD and chemical vapor deposition (CVD) processes, with growth rates above 1 monolayer per cycle observed at higher temperatures, as shown in Fig. 1. For a reduced growth temperature growth rates start to approach these expected for the ALD process. The growth rate as a function of growth temperature is constantly decreasing from 0.4 nm/cycle at 400 °C to 0.08nm/cycle at 280 °C [32].

High growth rates, but also electrical properties of ZnO films (high conductivity), may be in our opinion a very attractive property of zinc acetate, when used as a zinc precursor for ZnO deposition. Samples grown at temperature of 300 °C and above often showed metallic n-type conductivity, with temperature independent free electron concentration $n$ is in the range of $10^{20}$ cm$^{-3}$. We also observed that samples resistivity increases with increasing measurements temperature, due to enhanced scattering processes. This is behavior expected for a metallic type of conductivity. This property, together with a high transparency in the visible light region, makes such ZnO films suitable for applications as transparent conductive oxides (TCO films) in photovoltaics [33].

For samples grown at lower temperature typical semiconductor-like conductivity is observed. Free electron concentration drops down to $10^{16}$ cm$^3$ for films grown at a temperature below 300 °C [32]. Typical activation character of $n$ and resistivity is observed.

Unfortunately, use of the zinc acetate is not advantageous for a growth of uniform ZnTMO films. For ZnTMO an increased growth temperature results in non-uniform alloys. For such samples we commonly observed FM response and non-uniform SIMS profiles (of the type shown in the references [15,28]). Non-uniform Mn distribution was also concluded from the CL intensity maps, in the experiments discussed below.

A competition of the ALD and CVD growth processes results also in an increased roughness of films deposited at higher temperatures (see Fig. 1). ZnO films grown at an increased temperature with zinc acetate were fairly rough, with their RMS reaching 10 nm for films deposited at 400 °C. The XRD investigations indicate a lower quality of such samples. By a lower quality we mean here a mixed growth mode (see discussion in section 3.3), with the [1 0 .0] (*a*-axis) being the preferential growth direction at temperatures between 280 and 340 °C, and the [0 0 .2] growth direction (*c*-axis) being the preferential growth direction at higher temperatures (above 340 °C) [32]. SEM images for such disoriented sample are shown in Fig. 2.

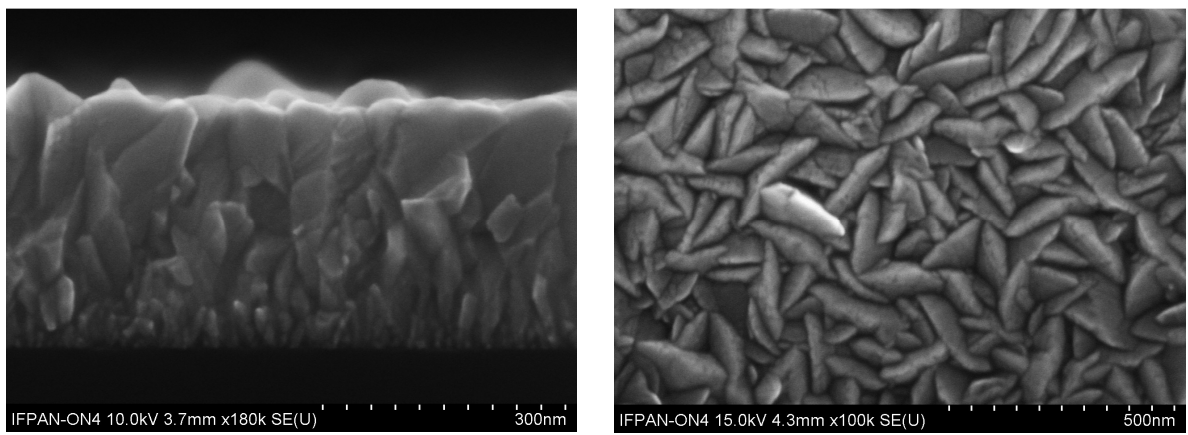

**Figure 2**: SEM images of disoriented sample with a mixed growth mode with most of ZnO columns (crystallites) oriented parallel to a substrate.

In further studies with the zinc acetate we slightly reduced the growth temperature to 300 - 360 °C, minimizing the contribution of the CVD processes. We underline here that when zinc acetate was used we always had a mixture of CVD and ALD processes and we could not clearly determine a so-called ALD growth window.

**3.1.2 Selection of Zn, Mn and Co precursors for ZnTMO deposition**

For deposition of ZnMnO (with zinc acetate as a zinc precursor, but also with DEZn and DMZn) we first used $Mn(thd)_3$ (2,2,6,6-tetra methyl -3,5-hepta nedione) as a manganese precursor. The temperature of precursors was as follows: 230 – 250 °C for zinc acetate (to avoid a direct decomposition of zinc acetate to ZnO), and 160-180 °C for $Mn(thd)_3$. Substrate temperature was 300 – 360 °C. Deposition and purging times were about ~0,5 s. Samples were grown on different substrates: on (0001)-orientated sapphire (for optical investigations), on glass (for electrical investigations), or on silicon (for magnetic investigations).

A microstructure of the so-obtained ZnMnO alloys was investigated in details. These investigations, and also magnetic ones, led to the following conclusion: even though we reduced substrates temperature to 300 ºC, this temperature was still too high to obtain uniform ZnMnO (we then found the similar situation for ZnCoO) alloys. Thus, in the further studies we replaced zinc acetate with DEZn and DMZn [24,28,30], which allowed further reduction of a growth temperature. We also replaced a manganese precursor. $Mn(thd)_3$ was replaced by $Mn(acac)_3$ (tris(2,4-pentanedionato) manganese(III)). This we done after the observation that use of the $Mn(thd)_3$ precursor promotes formation of inclusions of various Mn oxides (mainly $MnO_2$ and $Mn_3O_4$) to ZnMnO films. Temperature of the ALD process was reduced to 160 ºC, which turned out to be crucial for deposition of highly uniform ZnMnO films [28]. Low growth temperature, use of $Mn(acac)_3$ and of appropriate ratio of ZnO to MnO ALD cycles (see discussion below), finally allowed us to deposit uniform ZnMnO films, but also ZnMnO nanoparticles (grown with a hydrothermal process) with an uniform Mn distribution [29].

Experience we gained growing ZnMnO films we then used to deposit uniform ZnCoO layers. As a zinc precursor we used either dimethylzinc (DMZn) or diethylzinc (DEZn). As a cobalt precursor we used cobalt (II) acetylacetonate ($Co(acac)_2$), with deionized water used as an oxygen precursor. The precursors' temperature was as follows: DMZn - -30°C, DEZn – RT, $Co(acac)_2$ - 150°C, and deionized water - 30°C. Substrates (sapphire, GaAs, Si, glass, ITO-glass) temperature was 160°C, the one found optimal in the case of ZnMnO [28].

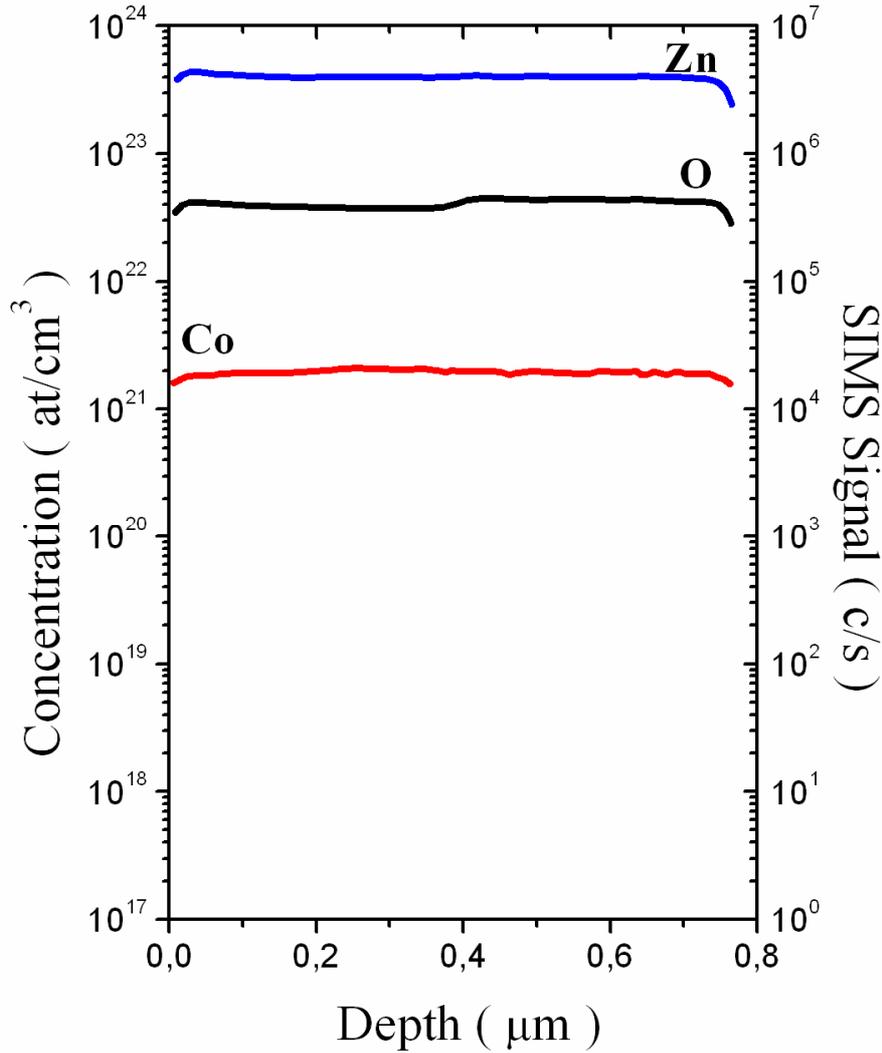

**Figure 3.** SIMS profiles of ZnCoO film grown by the ALD using DMZ as a zinc precursor, cobalt (II) acetylacetonate and water vapors. Deposition temperature was 160 °C. ZnCoO film was deposited with the 8:1 ratio of Zn-O to Co-O ALD cycles.

Their in-depth uniformity was investigated with the SIMS profiling, as shown in Fig. 3. In Fig. 3 we show the results for the ZnCoO sample grown using DMZ at 160 °C temperature using 8:1 ratio of the ALD cycles of Zn-O to Co-O (see discussion in the Section 3.5). ZnCoO (see Fig. 3) and ZnMnO samples grown at such conditions are uniform and show a paramagnetic response in the magnetic investigations. This observation strongly supports our initial assumption that FM response is a fingerprint of samples non-uniformity. We also observed that films grown with DMZn are more uniform that the ones grown with DEZn. However, due to difficulties in purchasing pure DMZn, most of the following investigations were performed with DEZn used as a zinc precursor.

### 3.1.3 Short summary on precursors selection

Summarizing, zinc acetate is precursor of choice if high growth rates and high free electron concentration are required. If reduction of a growth temperature is needed, then DEZn or DMZn precursors should be used. For ZnTMO alloys use of these two precursors is clearly advantageous. ZnTMO films obtained with these two precursors show uniform TM distribution. Surprisingly, we found that use of Mn(thd)$_3$ results in less uniform films, with inclusions of Mn oxides. Thus, we replaced this precursor with Mn (and also Co) acetylacetonate.

### 3.2 Growth modes of LT ZnO, ZnMnO and ZnCoO layers

All investigated ZnO and ZnTMO samples were polycrystalline and showed a columnar growth, with columns oriented along the *c*-axis, but with different orientations of *c*-axis versus a substrate (*c*-axis normal or parallel to a substrate). Four XRD peaks of ZnO in hexagonal phase ([10.0] (*a*-axis) for $2\theta = 31.770°$, [00.2] (*c*-axis) for $2\theta = 34.422°$, [10.1] for $2\theta = 36.253°$ and [11.0] for $2\theta = 56.603°$) were detected. Their relative intensity strongly depends on growth conditions. The [10.0] peak (the *a*-axis related) or the [00.2] (the *c*-axis related) often dominated the XRD spectra. We investigated relative intensities of these two XRD peaks. This allowed us to determine the preferential growth modes of our LT ZnO (ZnTMO) films.

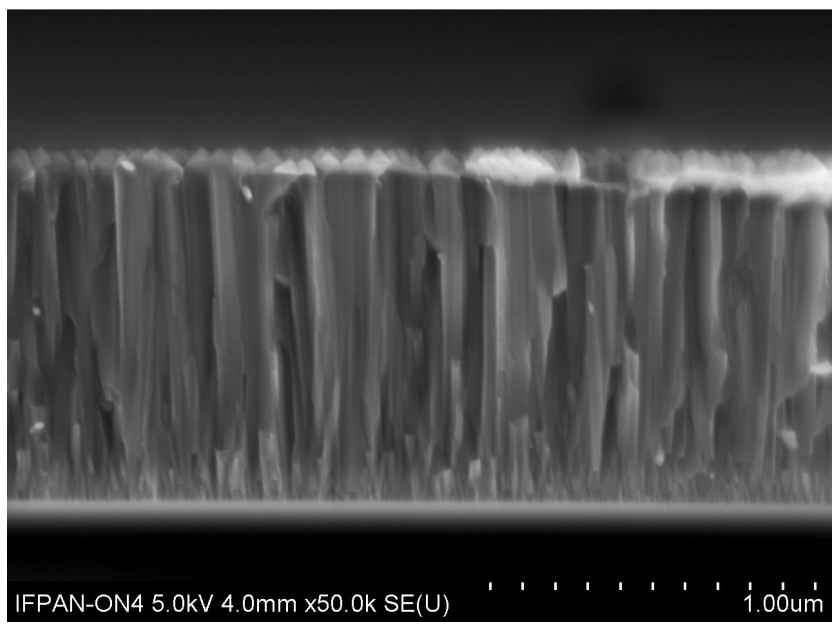

**Figure 4**: SEM images of a ZnO sample (grown with DEZn) with columns oriented normal to a substrate.

We observed three different orientations of our films, i.e., three different growth modes. ZnO, ZnTMO samples were grown with: 1) the *c*-axis normal to a substrate (see Fig. 4), 2) the *a*-axis normal to a substrate, and 3) with a mixed orientation of the columns (crystallites), with some of the columns normal and some parallel to a substrate (see Fig. 2). This classification is not very strict. In most of the cases we observed several XRD peaks (see Fig. 5), i.e., such samples should be classified as the ones with a mixed growth mode. Thus, by the *c*- or *a*-axis films we mean here the ones with their XRD spectra dominated by [00.2] or [10.0] axis, respectively. XRD peaks due to other orientations are either weak or not observed.

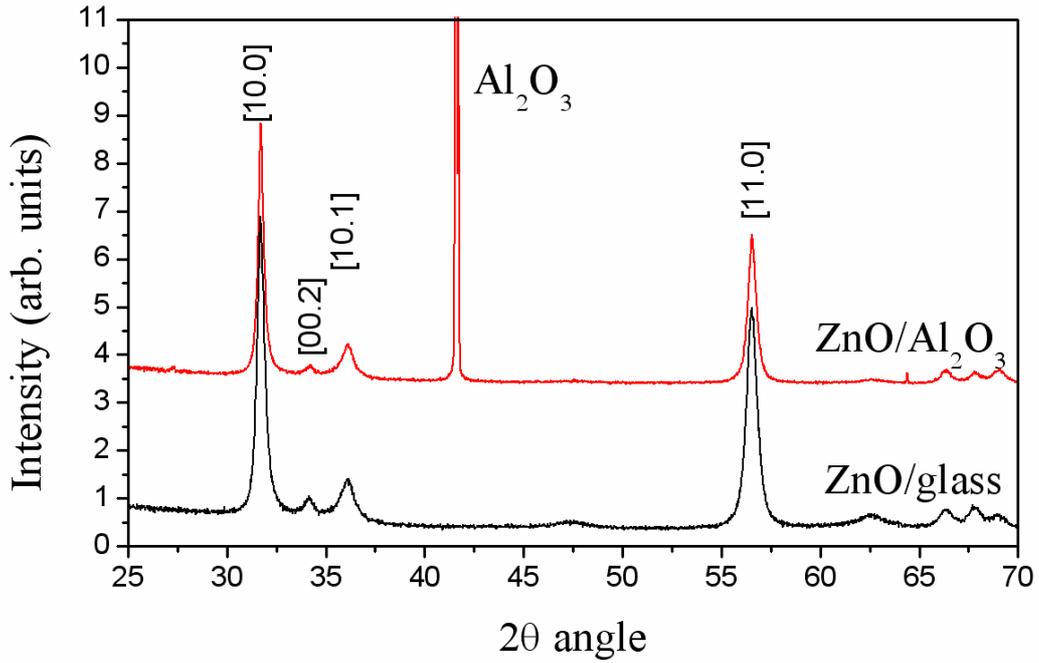

**Figure 5.** The XRD spectrum of ZnO sample grown by the ALD at 320 °C on c-sapphire and lime glass, using purging time of 0.9 s.

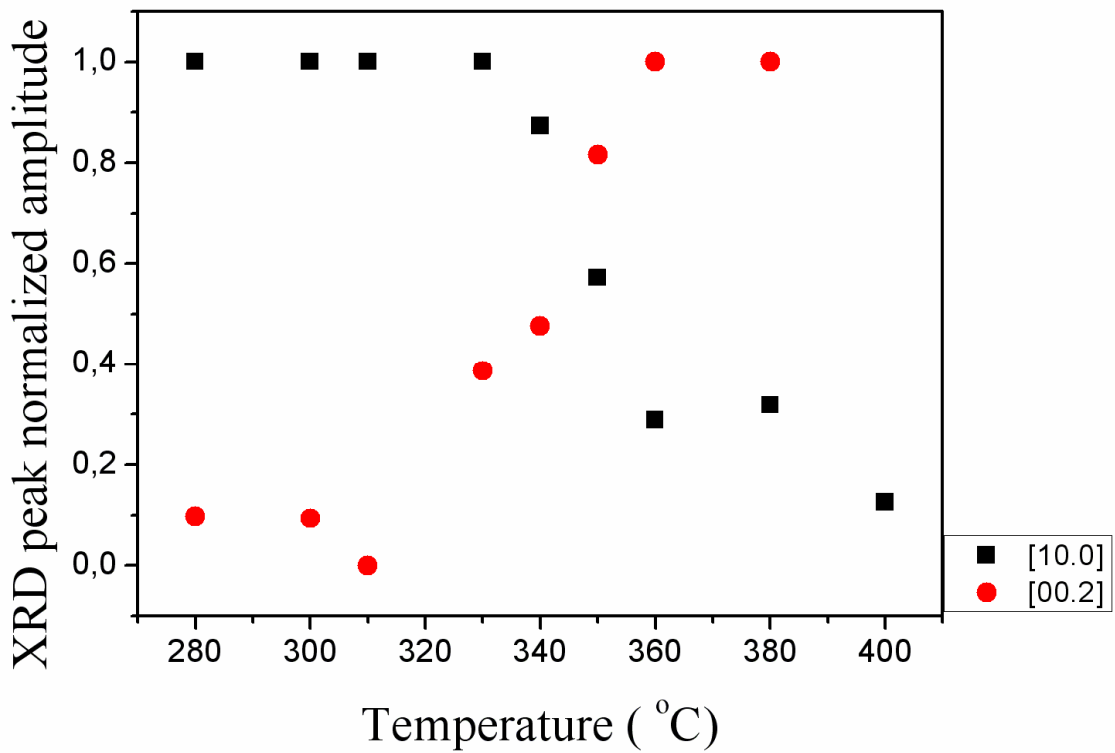

**Figure 6.** Relative intensity of the *a*-axis and *c*-axis related XRD peaks of LT ZnO films grown with zinc acetate and water vapor.

We first discuss growth modes of ZnO films grown with zinc acetate, based on the results of the XRD investigations, as the ones reported in the reference [32]. The XRD spectrum of the ALD-grown ZnO is presented in Fig. 5 for films grown at temperature of 320 °C, with a purging time of 0.9 s, and c-sapphire (ZnO/$Al_2O_3$) and lime glass used as substrates. The *a*-axis growth is observed for both types of substrates. By *a*-axis growth we mean here that [10.0] XRD peak dominates the XRD spectrum (see Figs. 5 and 6). Growth mode depends on growth temperature, as shown in Fig. 6.

Similar results we also obtained when zinc acetate precursor was replaced by DEZn or DMZn. The *a*-axis growth dominates for films grown at lower temperature [34]. For such films ZnO columns are deposited parallel to a substrate, but, as we concluded from SEM investigations, they are not aligned (as the ones shown in Fig. 2). In Fig. 4 we show cross-section SEM image for a film grown with columns normal to a substrate. The *c*-axis growth (*c*-axis normal to a substrate) is observed by us for ZnO and ZnMnO films grown with zinc acetate at increased temperature. In this case the XRD spectrum is dominated by the [00.2] peak, and two other are absent (as the one shown in Fig. 7 for ZnCoO films), or are very weak. The SEM images for a sample with a mixed growth mode are of the type shown in Fig. 2, with an interface region quite disoriented.

From the XRD investigations we concluded that the growth mode of ZnO and ZnTMO films depends on a temperature, on a sample composition, but also on a purging time, as discussed separately below. The latter observation was reported by us for the first time in the reference [32].

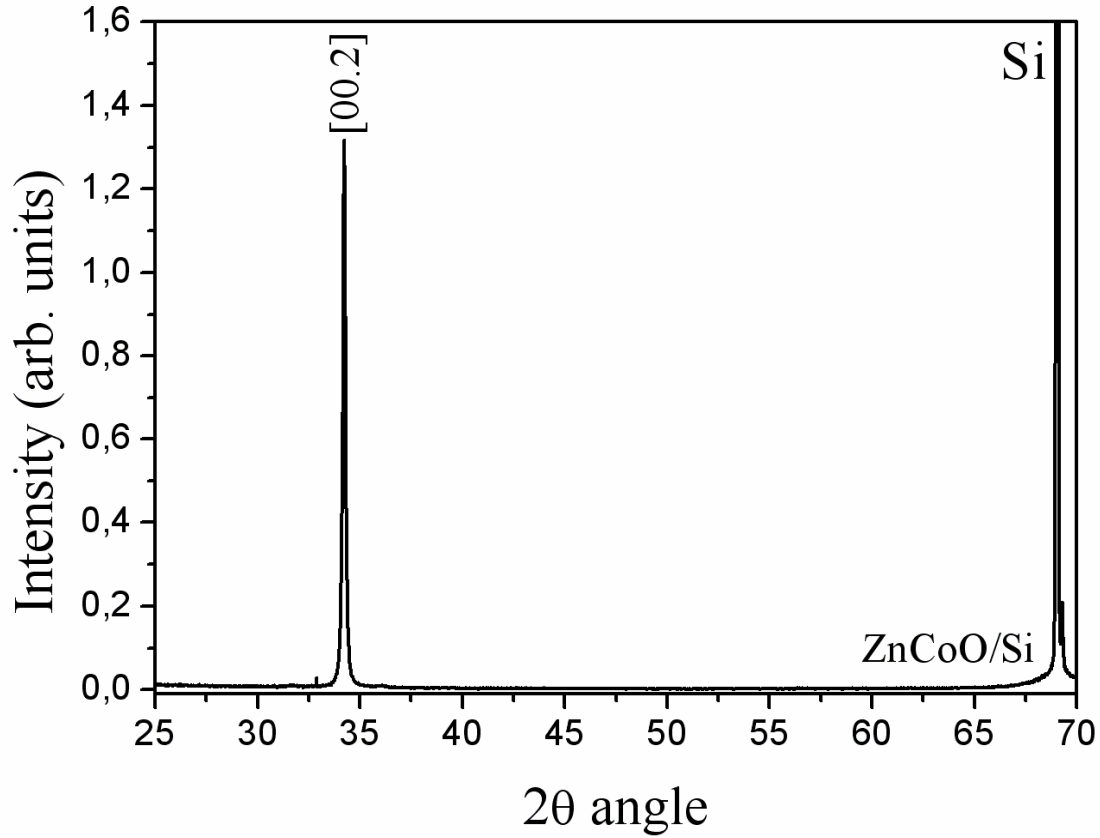

**Figure 7.** The XRD spectrum of ZnCoO film grown by the ALD at 160 °C using DMZ as a zinc precursor, purging time of 1.1s, and silicon as a substrate.

We performed several ALD processes to select the conditions to grow the films with the *c*-axis growth mode (i.e. with the *c*-axis normal to a substrate). In the case of ZnO layers grown with zinc acetate the [00.2] growth direction becomes preferential at temperatures above 350 °C. The *a*-axis ([10.0] XRD peak dominant) was the preferential growth direction at the temperatures between 280 °C - 340 °C (see Fig. 6). However, for zinc acetate we never got such clear *c*-axis growth, as the one shown in Fig. 7 for ZnCoO samples grown with DMZn. Similar results we obtained for the DEZn processes [34] and were reported for ZnO layers by V. Lujala et al. in the reference [35].

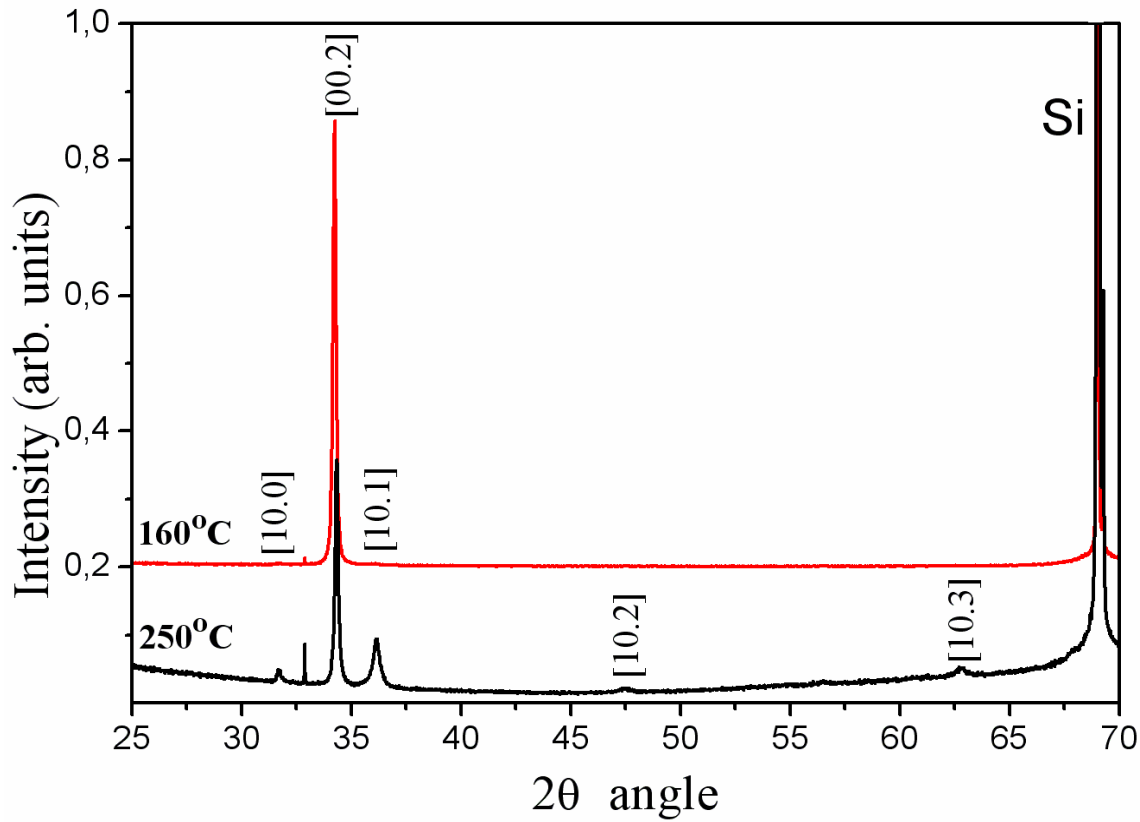

**Figure 8**. XRD spectra for ZnCoO films grown with DEZn at two different temperatures of: 160 °C and 250 °C.

The same conclusion about the dominance of *a*-axis and *c*-axis growth modes we also obtained from the XRD investigations of ZnMnO samples, but not of ZnCoO ones grown with DMZn. Whereas for ZnMnO the dominance of a given growth mode occurrs at the same conditions as for ZnO, a very different situation we found for ZnCoO films grown with DEZn and DMZn, as shown in Figs. 7 and 8. For DEZn/DMZn used as a zinc precursor we observed (see Figs. 7 and 8) that at a low growth temperature (below 200 $^o$C) the *c*-axis growth mode can be the preferential. The *c*-axis oriented growth of ZnCoO was replaced by a mixed growth mode at increased temperatures, as can be seen in Fig. 8. This observation, which has no clear explanation at present, shows how difficult is to get highly oriented films ones a low growth temperature is used.

For processes with ZnMnO and ZnCoO, in addition to the two growth modes discussed above – with *c*-axis normal to a substrate and *a*-axis normal to a substrate, we often observed the third one - a mixed mode (see Figs. 2 and 5).

Knowledge of the growth mode is not only academic. We found that this information is crucial to explain magnitude and anisotropy of the observed magnetization, as is discussed separately [30]. However, if low temperature processes of ZnO and ZnTMO deposition are required (to get uniform TM distribution, material suitable for hybrid structures with organic films, etc...), control of the growth mode is fairly difficult. It is more difficult in the case of zinc acetate and layers containing Co ions.

There is no clear explanation why the growth mode is so sensitive to growth parameters, such as temperature, precursor used, composition of alloys (Mn or Co), and finally a purging time (see next Section). Effects of a purging time are discussed in the following section.

**3.3 Influence of a purging time**

Growth modes depend not only on precursors used, process temperature, but also on other process parameters, In particular on a purging time [32]. Selection of a purging time affects samples orientation. Surprisingly, since we haven't found any correlation between pulse time of zinc acetate and orientation of the grown films [32].

Influence of a purging time we observed in the ALD processes with zinc acetate used as a zinc precursor. In a series of growth processes we changed a purging time (with nitrogen gas) from 1.1s up to 5.5s with a step of 1.1s. The same purging times were used after zinc and oxygen precursor pulses. As a growth temperature we selected 320 °C or 340 °C, the one at which *a*-axis related XRD peak dominates the XRD spectrum (see Figs. 5 and 6). The results are shown in Fig. 9.

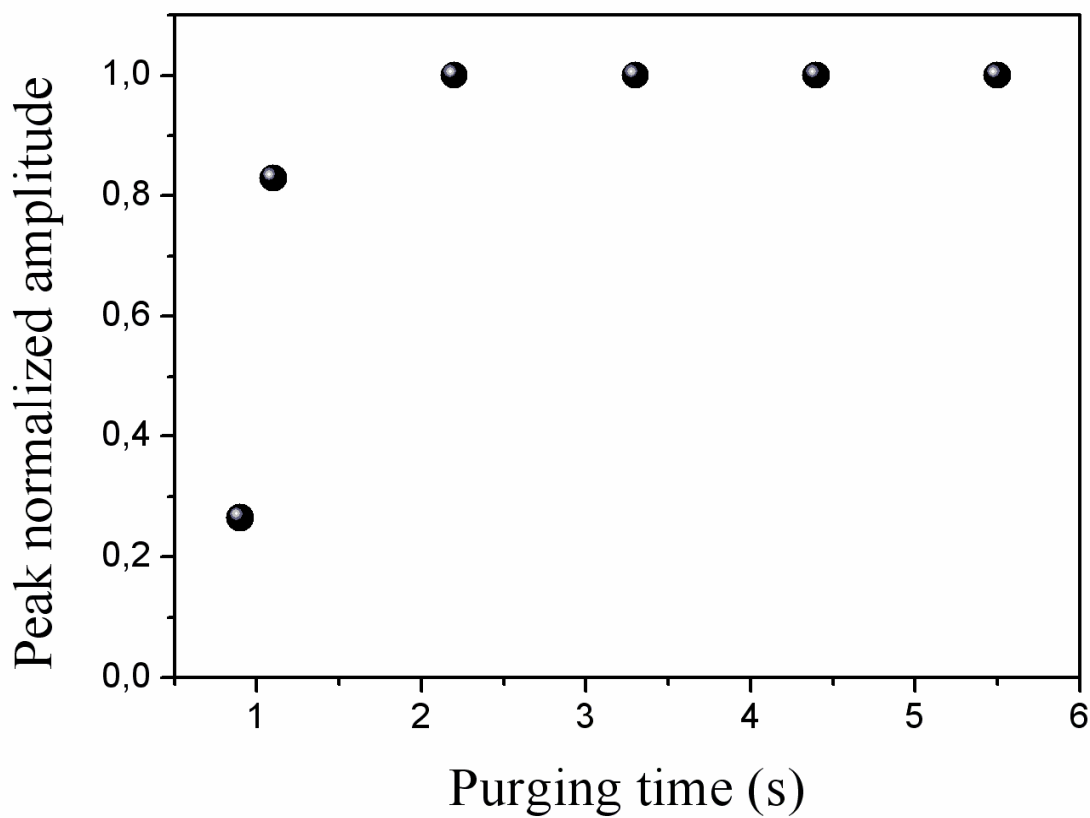

**Figure 9.** Intensity of the *c*-axis related XRD peak for ZnO films grown by the ALD at 340 °C using zinc acetate and different purging times. Figure shows a normalized amplitude of the [00.2] XRD peak.

We normalized the data shown in Fig. 9 in the following way – by 1 we mean the magnitude of the dominant XRD peak. At first, the *a*-axis related XRD peak dominates and the one due to *c*-axis is weak. This is situation similar to the one shown in Fig. 5. if longer purging times were used we observed a change of the dominant growth mode - from the one with the [10.0] axis normal to a substrate, to the one with the [00.2] axis (*c*-axis) normal to a substrate. For purging time longer than 2.2s the *c*-axis growth mode becomes dominant. Finally, only the [00.2] XRD peak was detected [32]. Surprisingly, this is observed without any effect on a growth rate of ZnO films.

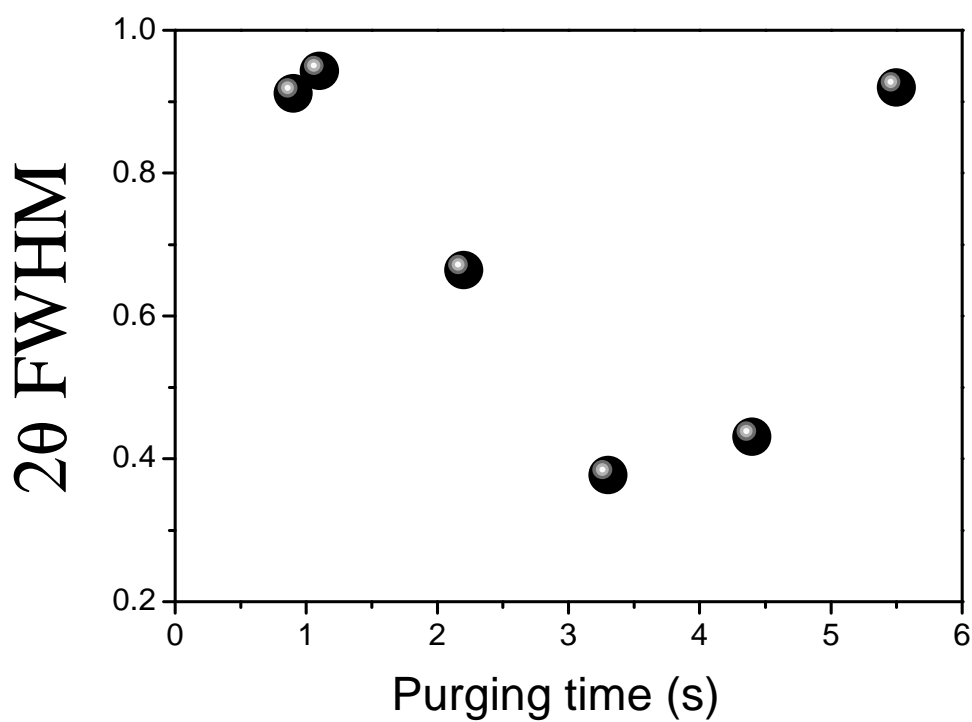

**Figure 10.** FWHM of the *a*-axis related XRD peak for ZnO films grown by the ALD at 340 °C using zinc acetate and different purging times. Similar dependence was observed for the *c*-axis related XRD peak.

Moreover, for processes with zinc acetate the purging time influences also the crystallographic properties of the films [32]. This we deduced from the analysis of FWHM of the XRD peaks (see Fig. 10), but also from the scanning and transmission electron microscopy investigations. With an increasing purging time FWHM of XRD peaks is reduced and reaches minimum at purging time of 3.3 s. Then, for longer purging times, it starts to increase [32].

From the FWHM of the relevant XRD peaks we estimated (using the Sherrer formula) effects of a purging time on a size of ZnO crystallites. Since the position and FWHM of a given XRD peak can also depend on strain conditions, defects, etc... we first checked if *a* and *c* lattice constants are equal to the ones of bulk ZnO. This we observed for our films grown with zinc acetate and for a growth temperatures above 320 °C. With an increasing purging time average size of crystallites rises from about 5 nm to about 15 nm.

Similar investigations we performed for ZnTMO layers. We found that the resulting uniformity of ZnMnO (ZnCoO) films depends on a length of precursor pulses and purging time. For example, for 1 s pulsing times of Zn (in the zinc acetate processes) and Mn precursors, the SIMS profile showed depth fluctuations of the Mn concentration. Then, a post growth annealing was necessary to get a uniform Mn distribution. Moreover, for these films (when grown with Mn(thd)$_3$) we still detected an unidentified high temperature FM response [14]. We then increased the purging time up to 4 s (from 1 s). A weak high-temperature FM phase was still detected in the films grown with Mn(thd)$_3$ at these conditions [14], but the FM response was much weaker. Origin of this effect is not clear. Thus, in the following processes we decreased the growth temperature to 280 °C in the zinc acetate processes and to 160 °C in the DEZn processes. Moreover, we changed ratio of the ALD cycles increasing number of Zn-O ones (10 to 1 or 9 to 1 zinc oxide to manganese oxide cycles). Such films were uniform. The magnetization investigations showed pure paramagnetic-type magnetic properties for these films.

## 3.4 Effects of the growth temperature and the identification of foreign phases

### 3.4.1 Temperature dependence of Mn oxides formation

There is a fundamental difficulty to obtain ZnTMO alloys, which are uniform and free of foreign phases. TM ions, such as Mn or Co, readily form various oxides. Inclusions of such Mn oxides are responsible for most of foreign phases in ZnMnO alloys. This may be a source of many incorrect reports on magnetic properties of ZnMnO, since all of these oxides have specific magnetic properties (see Table 1). This effect was investigated in details for manganese ions, as summarized in Tables 1 and 2.

**Table 1** Data on magnetic properties (magnetic response and critical temperature) of various Mn oxides (after [6]).

| Oxide | Magnetic order | $T_c$ |
|---|---|---|
| MnO | Antiferro- | 116K |
| $MnO_2$ | Antiferro- | 92K |
| $Mn_2O_3$ | Antiferro- | 76K |
| $Mn_3O_4$ | Ferro- or Ferri- | 43K lub 46K |
| $ZnMnO_3$ (cubic) | Spin-glass | 11-16K |
| $(Mn,Zn)Mn_2O_4$ (spinel) | Ferri- | ~30K |

**Table 2** Data on formation temperatures of various Mn oxides.

| Transition temperature | Oxide formation |
|---|---|
| 145 – 160 | $MnO_2$ |
| 530 | $MnO_2 \to MnO$ |
| 630 | $MnO_2 \to Mn_2O_3$ |
| 850 | $MnO_2 \to Mn_2O_3$ |
| 1100 | $MnO_2 \to Mn_3O_4$ |

In these tables we collected the data about formation temperatures of various Mn oxides and on their magnetic properties. In agreement with the data shown in the Table 1, MnO, $MnO_2$, and $Mn_2O_3$ show antiferromagnetic ordering, with the Neel temperature below RT (116 K, 92 K, and 76 K, respectively [6]). In the case of $Mn_3O_4$ situation is more complicated. Magnetic properties of $Mn_3O_4$ depend on a crystallographic structure of this oxide. $Mn_3O_4$ can be either ferromagnetic with $T_C = 43$ K, ferrimagnetic with $T_C = 46$ K (for a distorted spinel), or ferrimagnetic with $T_C = 40$ K (for a hausmannite form) [6].

Formation of various Mn oxides is temperature dependent, as reported in the reference [6] and is shown in Table 2. It was found that formation of most of the oxides is enhanced at increased temperatures. Thus, a low growth temperature is highly profitable, as already indicated by Sharma and co-workers [5], and by us for both ZnMnO [15,28,29] and ZnCoO [24,30]. This is why in further studies we lowered a growth temperature by replacing zinc acetate with DEZn or DMZn precursors. This finally allowed us to obtain inclusions free layers of ZnMnO (or the ones with a very low concentration of foreign phases), which showed only a paramagnetic response [15,28,29].

**3.4.2 Detection of inclusions of foreign phase**

The data collected in Tables 1 and 2 indicate that at growth temperature we used $MnO_2$ inclusions are the most likely, and that, in point of view of FM response, $Mn_3O_4$ and $(Mn,Zn)Mn_2O_4$ should be avoided. We thus employed several experimental methods to check if such oxides are present in our ZnMnO samples. Similar investigations were also performed for ZnCoO films.

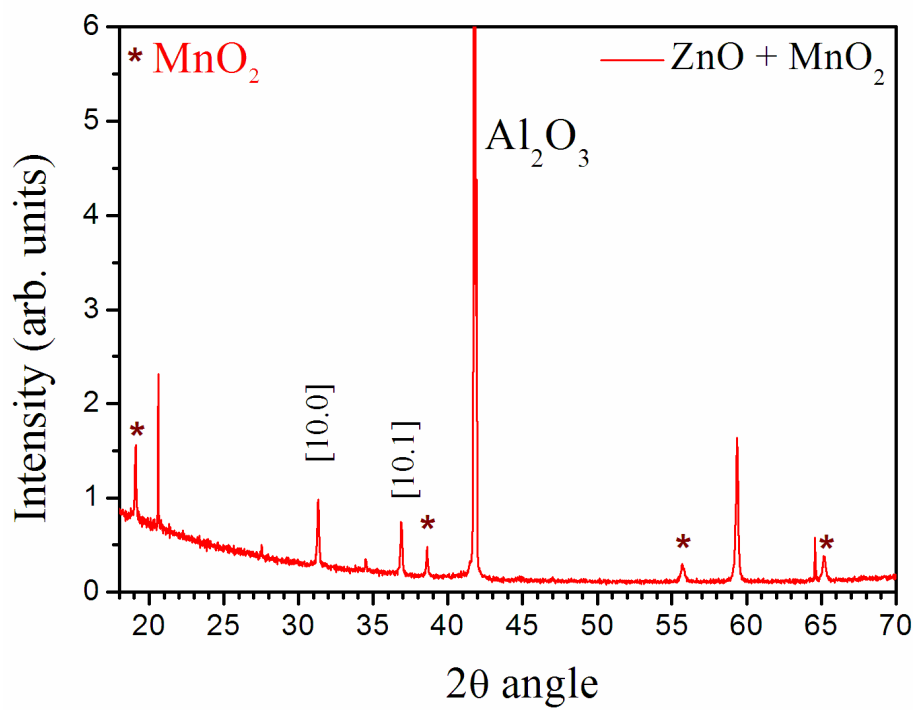

**Figure 11**. XRD analysis of ZnMnO films grown with DEZn and Mn(thd)$_3$. XRD peaks marked with stars indicate formation of $MnO_2$ inclusions.

Foreign phases in ZnMnO (due to inclusions of Mn oxides) were detected with the XRD for alloys grown with Mn(thd)$_3$ [14]. Since ZnMnO samples were grown at fairly low temperatures, the X-ray analysis of non-uniform samples indicates inclusions of $MnO_2$ only (see Fig. 11). This observation is in agreement with the data given in the Table 2. However, in most of other cases inclusions of foreign phases (due to TM oxides) were not detected with the XRD. Their concentration was too low for detection with the XRD. We thus employed X-ray absorption spectroscopy (XAS) to search for such inclusions, as discussed shortly below. For ZnCoO alloys the XAS study is discussed in details in the reference [30].

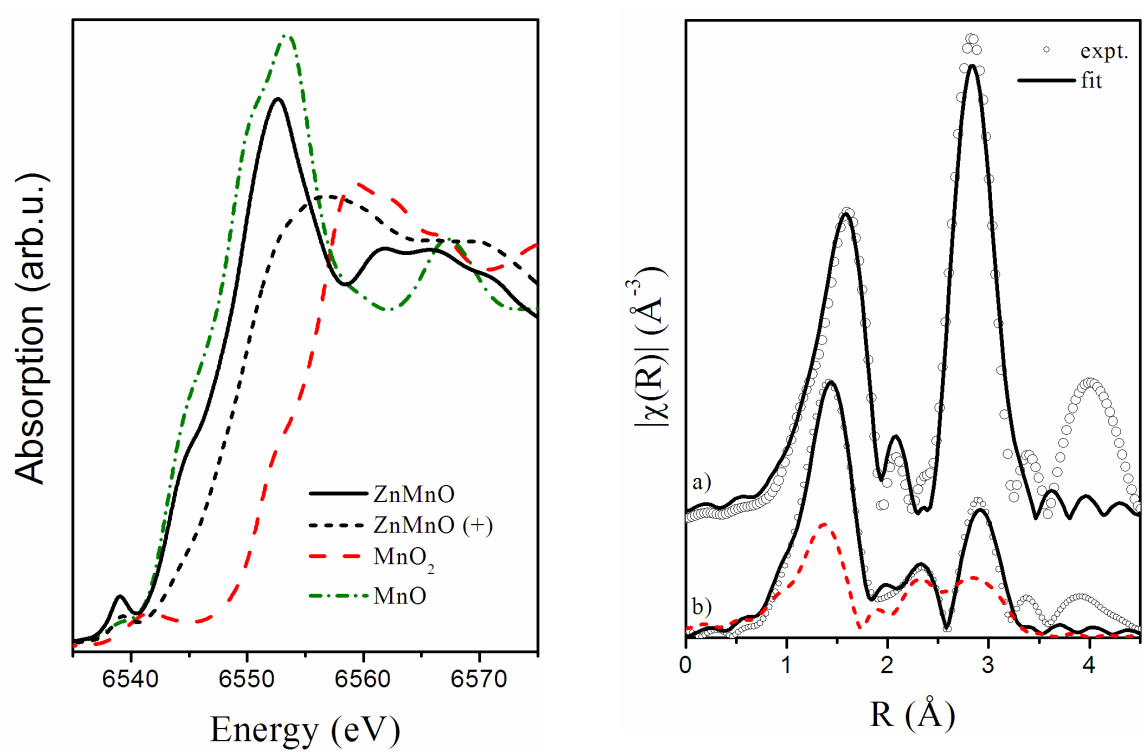

**Figure 12**. (left) The Mn K-edge positions of the XANES spectra of uniform (ZnMnO) and non-uniform (ZnMnO (+)) layers. The spectra are compared with the MnO and $MnO_2$ references. (right) The FT EXAFS spectra for: a) uniform ZnMnO; b) the layer with two phases: ZnMnO and $MnO_2$, where the contribution from $MnO_2$ is shown as a dashed line. Empty circles represent experimental results, full lines - obtained fits. Spectra are shifted vertically for clarity.

The XAS measurements (X-ray absorption near edge structure - XANES and extended X-ray absorption fine structure - EXAFS) were performed in fluorescence mode of detection at the K edges of Co and Mn. The measurements were performed at liquid nitrogen temperature to minimize the thermal disorder of the system. The reference samples of manganese oxides were measured in a transmission mode. EXAFS analysis was carried out using IFEFFIT data analysis package with the Athena and Artemis programs [36].

We employed the XAS technique due to its element selectivity and sensitivity, which makes it a perfect tool to investigate inclusions of foreign phases in the considered systems. In particular, the XANES spectra were analyzed, due to their sensitivity to the local structure around a selected element, which may give a fingerprint of the formed chemical bonds.

The results of the XAS experiment on the ZnMnO layers differ significantly for the uniform samples with a low Mn concentration (below 10%) and for non-uniform (ZnMnO (+)) with a high Mn content (above 20%). In Fig. 12 the K-Mn edge of two representative types of the ZnMnO layers are presented together with the MnO and $MnO_2$ references. The samples with a low Mn concentration are represented by the sample called ZnMnO and the samples with a high Mn concentration by the one called ZnMnO(+). It can be seen that the ZnMnO edge position is close to that of the MnO reference, which means that for this sample the ionicity of the Mn atoms is close to +2. The shapes of the considered spectra are different indicating, that the Mn atoms in the layer form the compound different from manganese oxides. In Fig. 12 (right) the Fourier transformed (FT) EXAFS spectra and their fits are shown. The crystallographic data for the ZnO structure were used for fitting of the results obtained for ZnMnO layers. The quality of the fit and the obtained parameters confirms that in this layer all Mn atoms are built into the ZnO structure forming a uniform ZnMnO layer.

In the case of the ZnMnO(+) sample, already the absorption edge shows distinctive differences (see Fig. 12 (left)). Its position is moved towards higher energy, which is as a result of an elevated average ionicity of the Mn atoms. This suggests the presence of two or more phases formed in the sample. Taking into account that the edge position of the $MnO_2$ reference sample is at higher energy, it is very likely that this compound exists in the sample.

The FT EXAFS spectrum of the ZnMnO(+) sample also shows the changes in respect to the ZnMnO layer (Fig. 12 (right)). The distance to the first coordination sphere (position of the first peak in the spectrum) is shorter and the magnitude of the second peak is much lower. Such changes indicate the coexistence of another phase. This hypothesis was confirmed by the EXAFS analysis. The fitting revealed that ~75% of the Mn atoms in the ZnMnO(+) layer is built into the ZnO structure forming ZnMnO alloy, while other ~25% are in form of the $MnO_2$ (or some other Mn oxides ($Mn_xO_y$)) inclusions. In Fig. 12 right (b) the contribution to the spectrum from the $MnO_2$ inclusions is shown as a dashed line. We conclude that the Mn atoms are quite easily incorporated into the ZnO structure up to some critical Mn concentration. Above this point the exceeding Mn atoms form manganese oxides.

Samples showing $Mn_xO_y$ inclusions were then analyzed with the ESR and SQUID magnetometry. In both investigations we observed a quite complicated magnetic response. The SQUID investigations identified $Mn_3O_4$ inclusions, not expected (see Table 2) due to a LT growth. We then found that $Mn_3O_4$ Mn oxide is already present in $Mn(thd)_3$ precursor. In the ESR study we detected three different Mn-related resonance spectra. At low temperature the ESR spectrum was dominated by a paramagnetic signal of "isolated" $Mn^{2+}$ ions, with a weakly resolved hyperfine structure of $Mn^{2+}$ (characteristic 6 lines ESR spectrum, due to $Mn^{2+}$ hyperfine structure). This signal was superimposed on a broad one, dominant at 100 K, attributed by us to a resonance of manganese in a spin-glass phase of strongly coupled antiferromagnetic $MnO_2$. Finally, above 200 K we observed a weak resonance, reflecting a contribution of a ferromagnetic phase. Origin of the latter signal is not known at present. Definitely, it is due to a presence of inclusions of some foreign phases in our films.

A complicated magnetic response of the films grown with $Mn(thd)_3$ motivated us to replace $Mn(thd)_3$ by another Mn precursor, i.e., to use $Mn(acac)_3$ as Mn precursor. In so-obtained films of ZnMnO (if grown at low temperature) inclusions of foreign phases are practically absent, as concluded from the ESR, SQUID investigations discussed in details in the reference [28]. For example, the ESR spectrum was dominated by a strong paramagnetic contribution due to manganese ions substituting Zn in ZnMnO. However, we could still find a very weak contribution due to inclusions of foreign phases. This we concluded from a difference between ESR signals measured at field cooling (FC) and zero-field cooling (ZFC) conditions. A small differential signal was obtained only for signals measured at temperatures below 50 K. SQUID investigations confirmed this observation. A weak FM response was detected, due to a foreign phase inclusions with a critical temperature $T_C$ of about 45 K, as the one reported for $Mn_3O_4$. From the SQUID investigations we estimate that one Mn ion per $10^4$ is present in the inclusions of $Mn_3O_4$ [14].

The data collected in the Tables 1 and 2 are for Mn oxides. Likely, similar oxides are formed for Co ions. Thus, we investigated ZnCoO samples with TEM, XAS (Fig. 13) and XPS [24,30]. There, as discussed in details in [30], we searched for inclusions of several Co oxides and Co metal in ZnCoO films. In ZnCoO even in the cases when inclusions Co oxides exist, no FM response was detected. We concluded that for ZnCoO alloys Co metal inclusions, and not Co oxides, are responsible for the FM response [24].

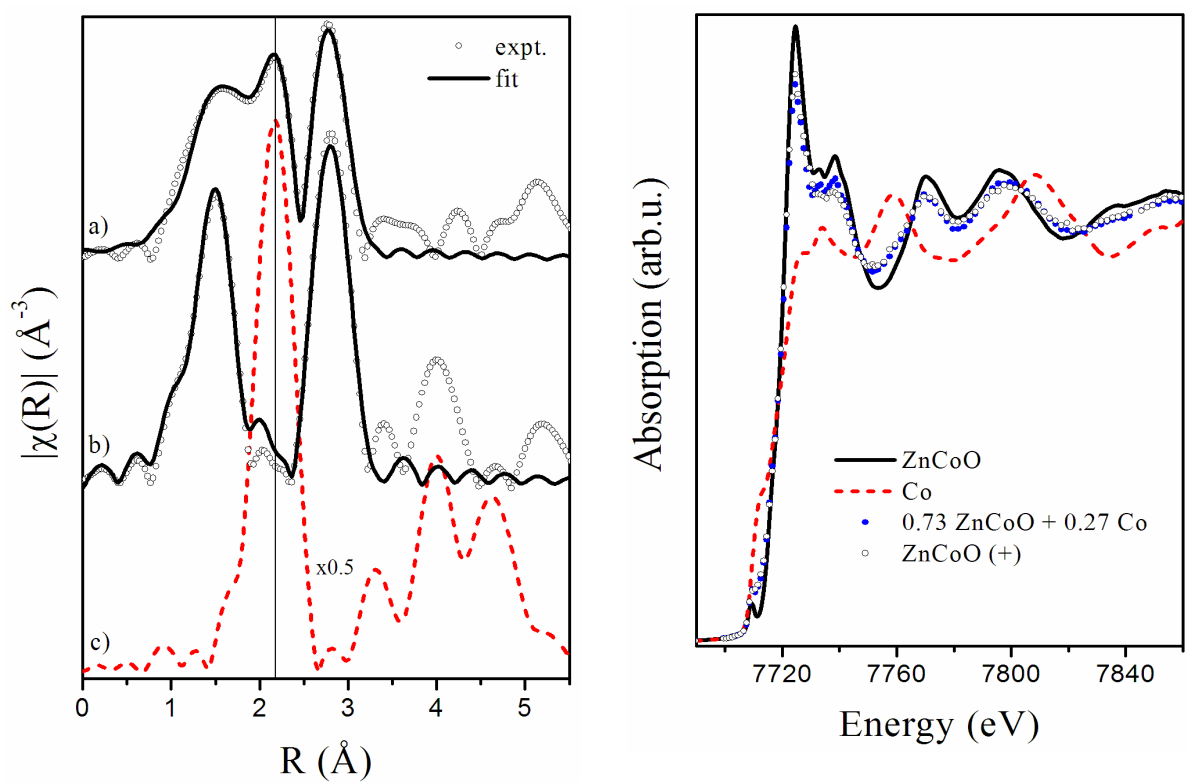

**Figure 13**. (left) The FT EXAFS spectra of: a) the ZnCoO(+) layer with two phases: ZnCoO and Co metal; b) the uniform ZnCoO layer. Empty circles represent experimental results, full lines - obtained fits. c) The Co metal reference spectrum (dashed line) is shown for the comparison. Spectra are shifted vertically for clarity. (right) Normalized XANES spectra at the Co K-edge for: the uniform ZnCoO sample (full line), the Co reference (dashed line), the weighted sum of them (full circles). Results for the ZnCoO(+) sample are indicated with empty circles.

XAS investigations were performed for ZnCoO films with different Co concentrations and samples uniformity. Contrary to the ZnMnO layers, there is no evident dependence on the Co concentration. The XAS investigations of ZnCoO layers show differences between the samples with uniform (ZnCoO) and non-uniform (ZnCoO+) Co distribution. Two types of spectra are distinguished.

In Fig. 13 (left) the FT EXAFS spectra for two representative layers ZnCoO and ZnCoO(+) are shown. For the uniform ZnCoO layer the performed fit (also shown in Fig. 13) confirms that all Co atoms are built into the ZnO lattice forming ZnCoO alloy. This is not the case for the ZnCoO(+) sample (Fig. 13). This sample was grown on purpose to be less uniform (see section on the ALD pulses). It can be easily noticed that FT EXAFS data are affected by contribution of another phase. An additional peak around 2.3 Å can be seen. This peak is consistent with the first sphere for the metallic Co, as shown in Fig. 13 (c). The EXAFS analysis of the ZnCoO(+) spectrum reveals that in this sample about ~73% of the Co atoms are built into the ZnO lattice, forming ZnCoO alloy, whereas about ~27% of Co atoms forms Co metal inclusions.

The same conclusion can be drawn from the XANES spectra. XANES spectrum of a selected element is a weighted sum of the spectra of all compounds containing such element in an investigated material. Therefore, adding the expected spectrum for 73% Co atoms in ZnCoO and 27% in metallic Co should result in a spectrum similar to the one observed by us for ZnCoO(+) sample. In Fig. 13 (right) we show the XANES spectra of the uniform ZnCoO sample and of the Co metal measured for the reference. We also show their weighted sum as well as the ZnCoO(+) spectrum. It can be seen that the signal of the ZnCoO(+) sample is very close to the weighted sum, which is consistent with the results of the EXAFS analysis. Further details of the EXAFS and XANES are given in the references 30 and 37.

The high-resolution depth-profiling XPS studies have been performed to determine the chemical state of cobalt in ZnCoO films. Prior to XPS measurements surface of films was sputter-cleaned by 500V Ar$^+$ ions. The XPS analysis was carried out at surface and within the bulk of a film in a crater created by subsequent sputtering steps. In Fig. 14 we show spectrum taken after removing of 15 nm of material. Sputtering was continued until an interface was reached. The representative Co 2p spectrum is shown in Fig. 14. Similar XPS spectra from a volume of the sample were observed for ferromagnetic and paramagnetic films. The Co $2p_{3/2}$ and Co $2p_{1/2}$ contributions are separated by about 15 eV and are observed at the binding energy (BE) 795 eV and 780 eV, respectively. The main Co 2p contributions are accompanied by the broad shake-up satellites detected at larger BE.

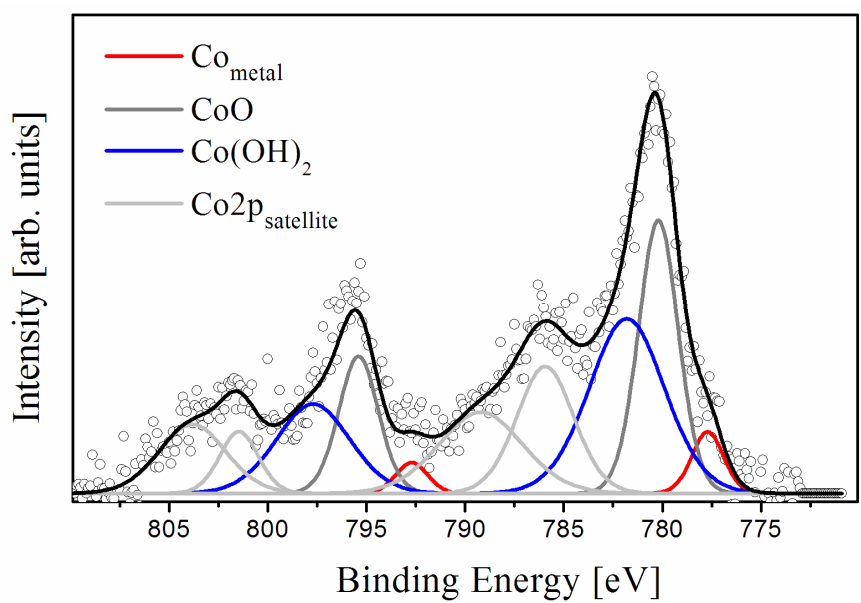

**Figure 14**. (color online) The Co2p core level XPS spectra of ZnCoO measured after removing 15 nm of the material by Ar$^+$ sputtering.

Deconvolution analysis of the Co2p$_{3/2}$ XPS spectrum was performed. Three forms of cobalt compounds can be distinguished inside ZnCoO samples. The strongest component, situated at BE equal to 780.3 eV, is due to cobalt oxide (Co substituting Zn in ZnO lattice). Strong satellite peaks located at BE between 784 eV and 791 eV confirm this identification. For the $Co_3O_4$ much weaker satellites are observed [38,39]. Another $Co2p_{3/2}$ XPS state, located at BE=781.8 eV, can be assigned to cobalt surrounded by –OH groups. This peak is accompanied by the strong satellite peak at BE about 786 eV. The $Co2p_{3/2}$ contribution at the BE=777.8 eV is due to a metallic cobalt [38]. This state was observed in both paramagnetic and ferromagnetic ZnCoO films, which indicates that some Co atoms in ZnCoO layers form metallic clusters and that these clusters are present both in ferromagnetic and paramagnetic films.

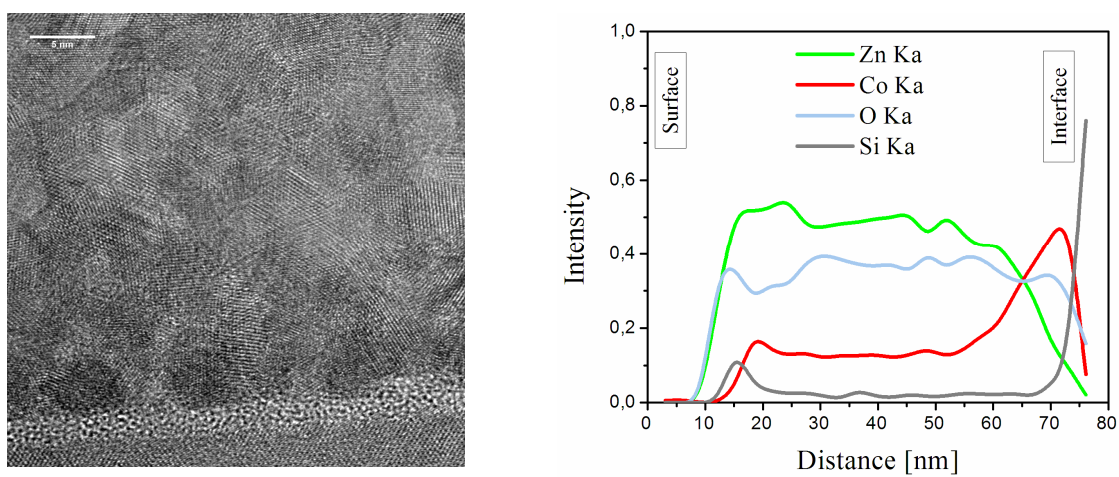

**Figure 15**. (a) TEM image of the ferromagnetic ZnCoO film. We observe metallic Co clusters with diameter of about 5 nm at the ZnCoO/Si surface. (b) (color online) The accumulation of cobalt observed in the EDX measurement taken at the cross-section of ZnCoO films.

From the TEM (Fig. 15) and XMCD (see reference [30]) investigations we concluded that the observed ferromagnetic response does not depend on the presence of metallic inclusions only, but it is related to formation of mesh-like structure of close lying and strongly coupled metallic cobalt nano-inclusions (clusters) at the ZnCoO/Si interface [30]. These clusters are of about 5 nm size and smaller and are situated very close each other as shown in the TEM image (Fig. 15 (a)). In Fig. 15 (b) the accumulation of cobalt at ZnCoO/Si interface is presented, as observed in EDX in HRTEM investigations. We concluded that Co metal inclusions at the interface region of the ZnCoO alloys are responsible for the FM response [24].

The results presented below allow selection of the appropriate ratio of Zn-O to Mn/Co-O ALD cycles to avoid formation of various Mn/Co oxides and Co metal inclusions.

### 3.5 Selection of the ALD cycles

We tested several sequences of the elementary ALD cycles to obtain uniform alloys of ZnMnO and ZnCoO. The films were grown as follows: we started with a given number of the ALD cycles to grow ZnO (zinc precursor – purging - oxygen precursor - purging). Then, a given number of Mn-O (Co-O) cycles (manganese (cobalt) precursor – purging - oxygen precursor - purging) was introduced. Ratio of these cycles was selected depending on the process temperature and the required Mn (Co) concentration.

The ratio of ZnO to MnO (CoO) cycles varied from 2:1 to 10:1 in various options. For example, a series of ZnCoO films was grown using 8:1, 80:10 and 80:5 ratios of these cycles [24,30]. Films with thick ZnO "spacers" (large number of ZnO ALD cycles) followed by 1-2 Co-O cycles were more uniform and in the consequence showed only a paramagnetic response (see e.g. discussion given in the reference [24,30]).

Regarding TM fraction in the alloys, we first assumed that to get a given value of x in $Zn_{1-x}TM_xO$ we should select an appropriate number of Zn-O to TM-O ALD cycles following each other. For example, to get x=0.1 nine Zn-O ALD cycles should be followed by one TM-O. Analogously, for x=0.5 one Zn-O ALD cycle should be followed by one TM-O. This assumption turned out to be incorrect, as will be discussed later on. We found that the resulting TM concentration in a "volume" of the sample is not exactly following the ratio of the ALD cycles.

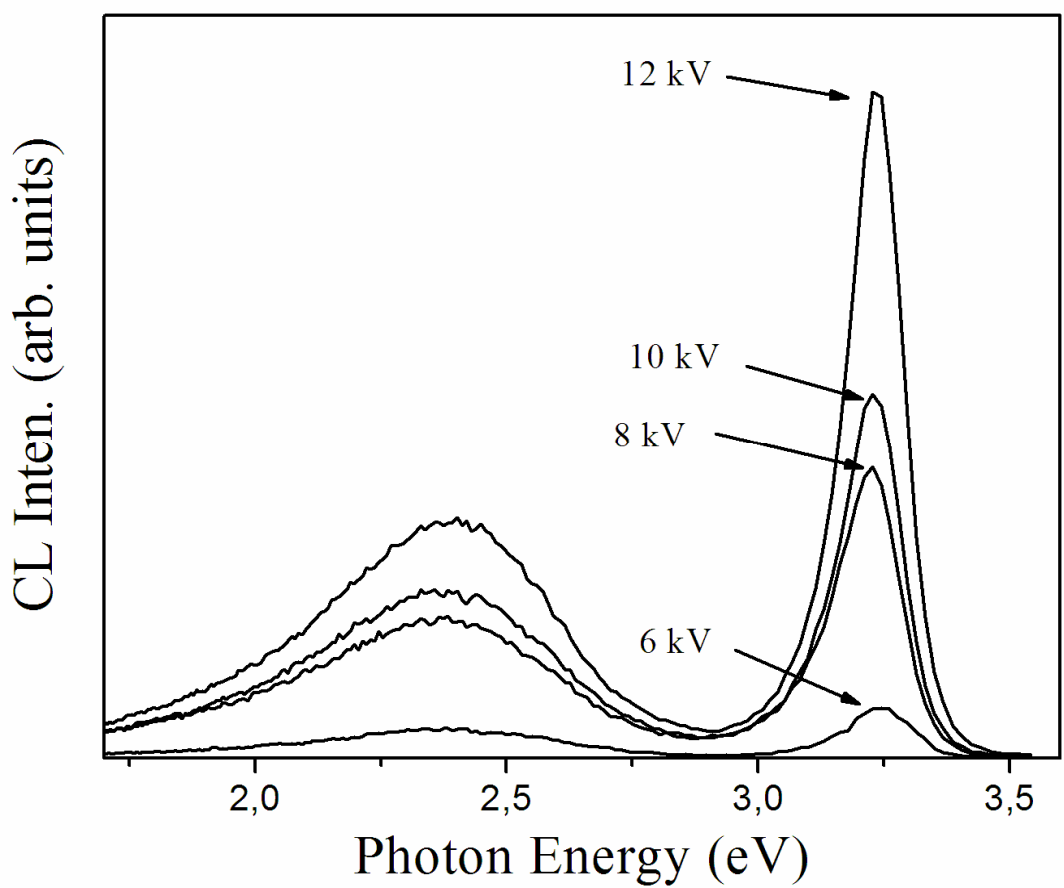

**Figure 16**. CL depth-profiling of a non-uniform ZnMnO film with manganese ions accumulated at the surface. CL is deactivated when excited from the Mn reach regions.

For ZnMnO alloys we first tested 2 to 1 ratio of the Zn-O to Mn-O ALD cycles. The resulting layers were highly in-depth and in-plane non-uniform, as we concluded from the SIMS, XPS and from in-plane maps of cathodoluminescence (CL) intensity, and from the CL depth-profiling [40,41]. In the CL investigations (see Figs. 16 and 17) we utilized the fact that both Mn and Co deactivate a visible emission of ZnO [42]. In the consequence, TM rich areas are observed as dark regions in the CL maps (see Fig. 17). Otherwise, in-plane variations of the CL intensity reflect columnar microstructure of our films observed in the SEM images (see Fig. 17). If TM distribution is not depth uniform, CL collected at different accelerating voltages, i.e., coming from different depth regions of the sample, (see Fig. 16) reflects this non-uniformity. For example, CL coming from the surface of the sample, where Mn accumulation was detected with SIMS, is the weakest (Fig. 16).

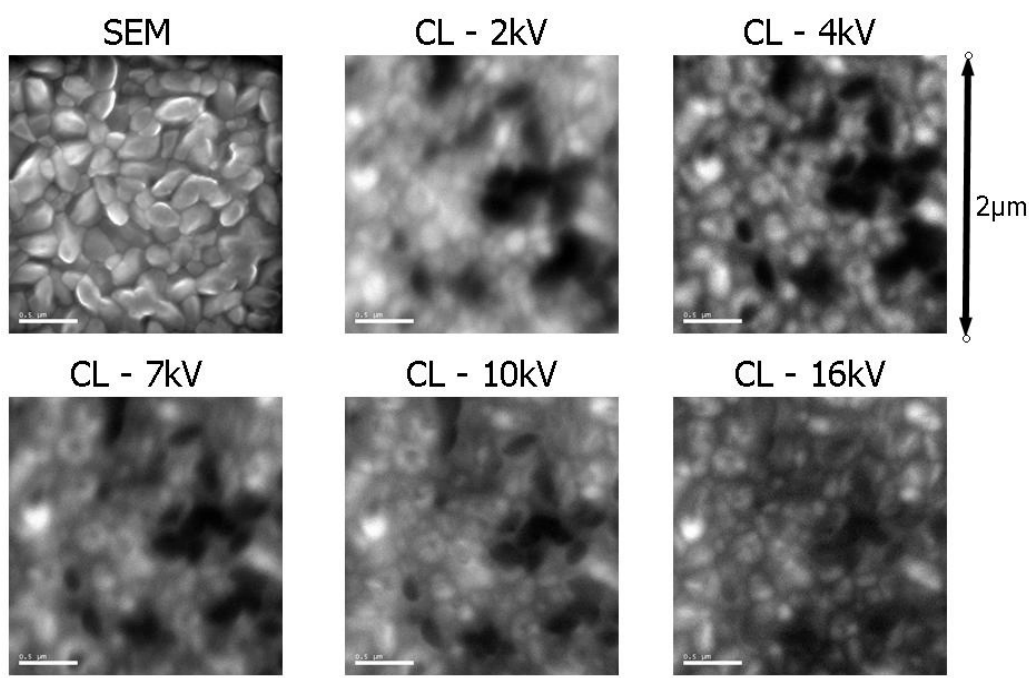

**Figure 17**. Maps of in-plane CL variations obtained from CL depth-profiling of 2x2 µm area. Investigations were performed for the ZnCoO film annealed at 800°C. The data were taken at accelerating voltages from 2 to 16 kV, i.e., at conditions in which signal is collected from its surface (at 2 kV), depth of the sample, and from the interface region (at 16 kV).

Most of the CL investigations was performed for ZnMnO alloys, since for ZnCoO CL was deactivated at fairly low Co fractions in samples. More uniform ZnMnO (ZnCoO) layers we obtained by separating the Mn-O (Co-O) ALD cycles (1 to 2) by up to 10 zinc oxide cycles.

For ZnMnO the uniform alloys were obtained for the sequence of 10 cycles of Zn-O (zinc and oxygen precursors applied sequentially 10 times) and up to 5 cycles of Mn-O (Mn and oxygen precursors applied sequentially 5 times) [28]. Large ZnO "spacer" accounts for a fast diffusion of Mn (and also Co, as discussed below). We observed that if the "spacer" is too thin, TM ions either float on a growth front or accumulate at interfaces. For example, for ZnMnO with a small ratio of the Zn-O to Mn-O cycles most of Mn ions out diffused to the surface of the film, as we deduced from the depth-profiling CL investigations [39,40]. Fig. 16 shows that for such samples CL intensity excited from the surface close area is the weakest.

Similar conclusion can be drawn from SIMS investigations, shown in Fig. 18 for ZnCoO alloys grown with different Zn-O to Co-O ALD cycles, ones keeping all other growth parameters the same.

We tested several possible ratios of the ALD cycles. Basically, similar ratio of the Zn-O to Co-O ALD cycles (as for ZnMnO) is required to get films with a paramagnetic response only [15,28,24,30]. When number of Zn-O ALD cycles was too small Co ions easily diffused through the layer and where floating on growth front (see Fig. 18). In turn, when number of subsequent Co-O was increased (e.g. to five or ten) we observed enhanced Co-Co coupling due to a non-uniform in-plane distribution of Co ions. This was observed even though SIMS profiles indicated quite uniform depth distribution of Co ions, as we show in Fig. 18 for selected ZnCoO alloys.

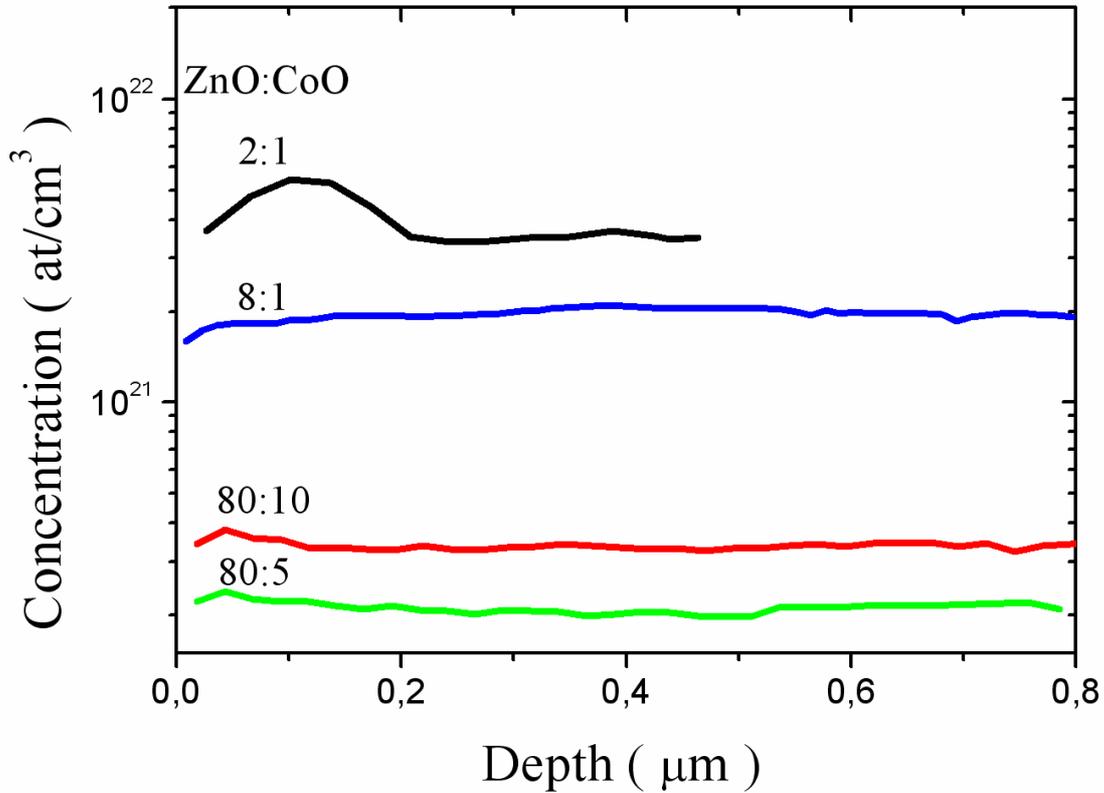

**Figure 18.** SIMS profiles of ZnCoO films grown by the ALD using different ratios of Zn-O to Co-O ALD cycles. Deposition temperature was 160 °C.

We also checked that 8:1 and 80:10 ratios are not fully equivalent. In point of view of a FM response the results seem to be similar. LT of the growth plus 8:1 or 80:10 ratio of the ALD cycles resulted in paramagnetic films at RT. However, films grown with 80:10 ratio show property of digital alloys with non-uniform TM distributions, which must be taken to account when magnetic properties of the samples are analyzed [30]. We observed that the antiferromagnetic (AF) Co-Co coupling is enhanced in the case of films grown with the subsequent 10 Co-O cycles.

### 3.6 Influence of layers thickness

A common feature of the ALD grown ZnO films on lattice mismatched substrates is that the layer structural quality improves with a thickness. This we observed in several experiments – in the XRD, SEM (see Figs. 2 and 4), TEM, CL investigations (see Figs. 16 and 19). For example, in the CL depth-profiling investigations we observed that interface close area is strongly defected [41]. Excitonic emission increases in intensity once the thickness of the layer is increased or when we moved the excitation point from the interface area (see Fig. 19).

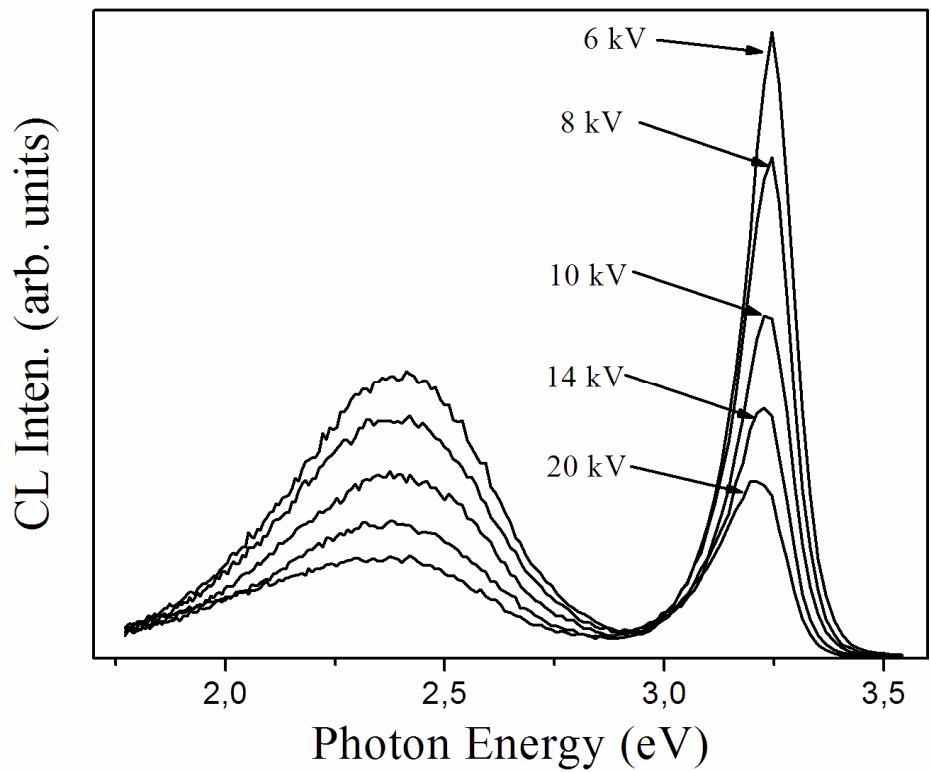

**Figure 19**. CL depth-profiling of ZnMnO film with an uniform Mn distribution.

In Fig. 19 we show results of CL profiling for ZnMnO film with depth uniform Mn distribution. Then, the changes of the CL intensity reflect a sample crystallographic quality.

Weak band edge emission is observed from the highly defected interface region. Defect related CL is also the weakest when excited from the interface region (i.e., at increased accelerating voltage). Intensity of emission rises once CL is excited from surface close areas.

An increased concentration of deep defects (of vacancies e.g.) affects uniformity of TM distribution in the ZnTMO alloys. This effect we studied in details for ZnCoO films, for which we observed that the FM response comes from the interface close layer of the samples [24]. The detail results we got from the XPS and TEM investigations. As already mentioned, we collected the XPS signal from the surface area of a given sample, then this layer was removed by sputtering and the XPS signal was recorded from the so-created crater. The process was repeated until the interface region was reached [24,30]. This study gave us the detail information on sample uniformity and on foreign phases present. For non-uniform samples we observed strong Co metal related XPS signal coming from the interface region. The XMCD investigations [24,30] confirmed that the FM response observed comes from this region of the sample and is due to Co metal inclusions.

**Conclusions**

ZnMnO and ZnCoO samples were grown by a range of possible growth methods – by pulsed laser deposition [12,16], molecular beam epitaxy [43], peroxide molecular beam epitaxy [44], MOCVD technique [45], magnetron sputtering [46], direct current reactive magnetron co-sputtering [47], sol-gel [48], and many other methods. Present study shows advantageous properties of the ALD growth method. By selecting an appropriate growth temperature and ratio of Zn-O to TM-O ALD cycles we could deposit uniform ZnTMO alloys. Also selection of zinc and TM precursors turned out to be important, as we found in the case of ZnMnO layers. Magnetic investigations of uniform layers gave a paramagnetic response. For films grown deliberately non-uniform we got a complicated magnetic response. Such control of films properties allowed us to account for many of conflicting reports on magnetic properties of ZnTMO alloys.

The present results we cannot compare with other done on samples grown with the ALD. The present study is the first one performed on ZnTMO samples. We could only verify validity of our conclusion on important role of a low temperature growth. This we done for ZnMnO [29] and ZnCoO nanoparticles (not published results). Nanoparticles were grown by a hydrothermal method. We used similar precursors, as used in our ALD study. For these nanoparticles we found that samples grown at low temperature are paramagnetic. The one obtained in a high temperature process showed a FM response.

**Acknowledgments**

„Microchemistry" F-120 ALE reactor used by us was financed by Foundation for the Polish Science within the "Sezam" program. The research was partly supported by the European Union within European Regional Development Fund, through the grant of the Innovative Economy (POIG.01.01.02-00-008/08). ZnCoO investigations were partly supported by the European Research Council through the FunDMS Advanced Grant (#227690) within the "Ideas" 7th Framework Programme of the EC.